**On Kerr's corpuscle**

A. LOINGER

Dipartimento di Fisica, Università di Milano

Via Celoria, 16, 20133 Milano, Italy

ABSTRACT. — Simple changes of the radial co-ordinate deprive Kerr's spinning corpuscle of its marvellous properties.

As it is well known, Kerr discovered in 1963 the following solution of Einstein field equations [1]:

$$(1) \quad ds^2 = -\frac{\Delta}{\rho^2}\left(dt - a\sin^2\theta\, d\varphi\right)^2 + \frac{\sin^2\theta}{\rho^2}\left[(r^2 + a^2)d\varphi - a\,dt\right]^2 +$$

$$+ \frac{\rho^2}{\Delta}dr^2 + \rho^2 d\theta^2 \quad ; \quad (G=c=1),$$

where: $\rho^2 \equiv r^2 + a^2\cos^2\theta$ ; $\Delta \equiv r^2 + a^2 - 2Mr$ ; $a$ and $M$ are nonnegative parameters. *The case $M \geq a$ is of remarkable interest*. If $a=0$, eq. (1) gives the *standard* form of the solution (Hilbert, Droste, Weyl) of Schwarzschild's problem of the Einstein field of a mass point at rest (**not** the **original** form of solution obtained by Schwarzschild [2]).

Since the radial co-ordinate $r$ is arbitrary (see e.g. [3]), in lieu of the $r$ of eq. (1) we can choose any *regular* function of it. Thus, let us perform the following substitution:

$$(2) \quad r \rightarrow \left[r^3 + (2M)^3\right]^{1/3} \quad , \quad 0 \leq r < +\infty.$$

(*N.B.*: *Both* the *r*'s of this formula − and of formula (6), *vide infra* − go from zero to infinite.) We call the second member of (2) *Schwarzschild's expression*, because it is suggested by the **original** formula of this Author concerning the gravitational field of a material point [2]. Then:

$$(3) \qquad \Delta \equiv \left(r - M - \sqrt{M^2 - a^2}\right)\left(r - M + \sqrt{M^2 - a^2}\right) \to$$

$$\to \left\{ \left[r^3 + (2M)^3\right]^{1/3} - M \right\}^2 - M^2 + a^2 \; ;$$

accordingly, for this second form of $\Delta$ we have:

$$(4) \qquad \Delta \geq 0 \; ; \qquad \Delta\,[r{=}0] = a^2 \; ,$$

and

$$(5) \qquad -\Delta + a^2 \sin^2\theta = \left\{ \left[r^3 + (2M)^3\right]^{1/3} - M \right\}^2 - a^2\cos^2\theta + M^2 < 0 \; .$$

It follows immediately from (4) and (5) that − owing to substitution (2) − we have eliminated the event horizons, the stationary-limit surface, and the so-called ergo-sphere, which are inherent in the original form (1). No magic occurrence is now possible [4]. Kerr's spinning particle does not possess in reality any marvellous property. (Remark that the substitution of (2) in eq. (1) gives an expression for the d$s^2$ which coincides for *a*=0 with the d$s^2$ of Schwarzschild's paper [2]).

We can re-obtain the above result with many other substitutions, e.g. with the simple substitution

$$(6) \qquad r \to r + 2M \; , \qquad 0 \leq r < +\infty \quad ;$$

now:

$$(7) \qquad \Delta \to \left(r + M - \sqrt{M^2 - a^2}\right)\left(r + M + \sqrt{M^2 - a^2}\right) \; ;$$

therefore

(8) $$\Delta > 0 \; ; \quad \Delta [r=0] = a^2 \quad ,$$

(9) $$-\Delta + a^2 \sin^2 \vartheta = -r^2 - 2Mr - a^2 \cos^2 \vartheta < 0 \; .$$

Again, any amazing property of Kerr's object has disappeared.

***An obvious conclusion***: The form (1) of Kerr's metric has a *physical* meaning only for those values of the radial co-ordinate which are *greater than* 2*M*. Here the pretension to a geodesic completeness is a nonsense. Of course, the areas of the surfaces $r = M \pm \sqrt{M^2 - a^2}$ and of the stationary-limit surface $r = M + \sqrt{M^2 - a^2 \cos^2 \theta}$ are invariant expressions, but this is only a mathematical outcome, devoid of physical importance, because *the original system of co-ordinates does **not** possess any privilege*.

APPENDIX

When *a*>*M* there is no event horizon for Kerr's metric (1). We have an analogous situation in the problem of the Einstein field of a mass point endowed with an electric charge *q*. Indeed, the well-known Reissner-Weyl-Nordström-Jeffery solution can be written as follows:

(A.1) $$ds^2 = -\frac{\Gamma}{r^2}dt^2 + \frac{r^2}{\Gamma}dr^2 + r^2 d\theta^2 + r^2 \sin^2\theta \, d\varphi^2 \; ,$$

where

(A.2) $$\Gamma \equiv r^2 - 2Mr + q^2 \equiv \left( r - M - \sqrt{M^2 - q^2} \right) \left( r - M + \sqrt{M^2 - q^2} \right) \; .$$

For an *electron* *q*>*M*, and therefore the metric is quite regular, except at *r*=0.

Of course, substitutions like (2), resp. (6), give a $ds^2$ devoid of any event horizon.

*————————————*